# Full Mesh Dynamic Four-User Entanglement-based Switched QKD Network Integrated with SD-KMS

**N. Makris,[1,*] P. Konteli,[1] Z. Georgiadis,[1] A. Papageorgopoulos,[1] G. Anastasiou,[1] P. Papanastasiou,[1] K. Tsimvrakidis,[1] and G. T. Kanellos[1]**

[1]*Department of Informatics and Telecommunications, National and Kapodistrian University of Athens, Ilisia, 16122, Athens, Greece*

*nikmakris@uoa.di.gr

**Abstract:** We demonstrate a reconfigurable, four-user, full-mesh quantum key distribution network using polarization entanglement, integrated with a software-defined key management system to orchestrate dynamic optical switching and real-time key provisioning for quantum-safe applications.

## 1. Introduction

As quantum technologies advance, multi-user, reconfigurable quantum networks are becoming an increasingly active area of research with several experimental demonstrations worldwide. Entanglement distribution is considered the cornerstone of current and future quantum implementations with applications in quantum secure communications, i.e., Quantum Key Distribution (QKD), distributed computing and sensing [1]. Recent works have studied different quantum network architectures, demonstrating the potential to enable efficient and scalable distribution of entangled states in large-scale networks, proving its efficiency and scalability [1], [2]. In [1], a programmable 8 × 8 multi-port device was used to implement a global multiplexed quantum interconnected network of eight users by connecting two local, four-user multiplexed entanglement networks, while in [2] state multiplexing enabled a fully connected four-user network using six wavelength channels. Regarding entanglement-based QKD networks, notable demonstrations have explored multi-user connectivity, wavelength division multiplexing (WDM) and reconfigurable optical routing [3]-[8] .[3] was one of the first multi-user demonstrations showcasing four users in a fully connected graph without trusted-nodes, while [4] demonstrated a notable 40-user fully connected entanglement-based QKD. Approaches that do not rely on static architectures typically support dynamic reconfigurability in the network by employing wavelength selective switches (WSS) or low loss optical switches (LLOS) [5]-[8]. In [5] active optical switching was demonstrated in a four-user quantum network, enabling dynamic entanglement distribution between different user pairs while the works in [6] and [7] utilized WSSs to improve network flexibility and resource allocation. While these advancements pave the way for future multi-user QKD networks, prior work has not yet demonstrated their deployment within a fully stacked environment that supports key management systems (KMSs), dynamic network reconfiguration through optical switching, and real-time key consumption by system application entities (SAEs) all-together. In [8] the authors presented a complete stack by enabling dynamic network-topology reconfiguration in a six-user entanglement network by deploying a reconfigurable optical add-drop multiplexer (q-ROADM) comprising demultiplexers (DEMUXs), optical switches OFSs, and MUX/WSSs at the expense of additional optical losses, a critical parameter for the efficient distribution of entanglement. In this direction, [9] proposed a universal photonic switch designed to handle quantum information exhibiting low losses, highlighting the potential of switching technologies for scalable quantum networks. However, the work focuses on the switching technology rather than an end-to-end QKD network implementation.

In this work, we present a four-user, full-stack QKD network that integrates physical WDM entanglement distribution with dynamic reconfiguration through a minimal, loss-optimized optical switching scheme that is complemented with advanced dynamic key management and overarching network orchestration. The switching engine is simplified to relax the hardware and control complexity requirements while by using a custom-developed Software-Defined KMS (SD-KMS), this architecture bridges physical-layer entanglement switching with higher-layer key management to provide a unified framework for reconfigurable, multi-user QKD. The network implementation includes software-based post-processing producing secret key material at each node while the controller communicated directly with the LLOS to coordinate network reconfiguration. The network stack operated continuously with periodic switching events across successive duty cycles to enable full-mesh communication among all users. Simultaneously, the SD-KMS monitored and managed the network in real time, delivering quantum-generated keys to L3 encryptors for quantum-safe applications, such as IPsec VPN tunnels.

## 2. Experimental Setup and Implementation

In the experimental setup a commercial broadband type-0 polarization-entangled photon pair source (EPPS) was employed for the generation of entangled photon pairs across multiple wavelengths, centered at 1550.12 nm (ITU channel C34). Due to energy conservation during the spontaneous parametric down-conversion process in the

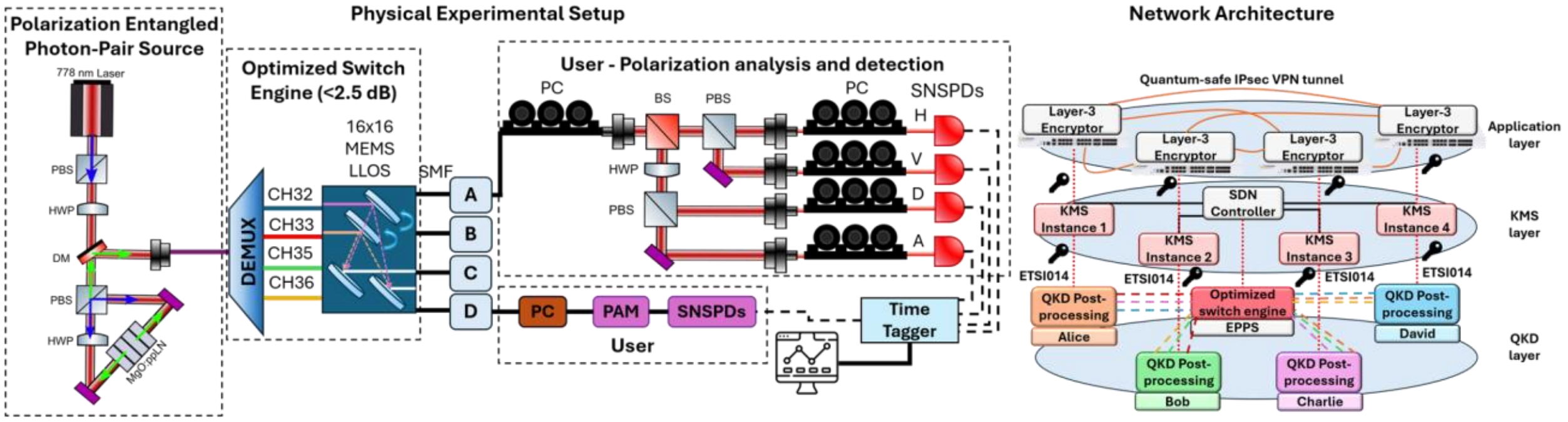

Fig. 1 The physical experimental setup of the optically switched four-user entanglement network, and the active QKD network layered architecture, depicting the fully meshed four QKD nodes topology, the KMS layer and the application layer where the QKD keys are consumed.

Magnesium-oxide doped periodically poled Lithium Niobate (MgO:ppLN) crystal within the source, entangled photons in the $|\Phi^{+}\rangle = \frac{1}{\sqrt{2}}(|HH\rangle + |VV\rangle)$ Bell state are generated only at correlated wavelengths. Correlated idler and signal photons were divided by utilizing a Wavelength Division Demultiplexer with 100 GHz channel spacing, while the dynamic distribution of the entangled photons between the users was carried out through a 16x16 Micro-Electro-Mechanical Systems (MEMS) LLOS. Each entangled photon pair corresponds to one logical link between two users. Each photon was guided in an input port of the LLOS, while each output was connected to a user analysis and detection module, as shown in Fig. 1. In this four-user actively switched network demonstration, two entangled photon pairs were used to establish simultaneously two links, alternating correlated photons among the users. The EPPS was connected to the users through the loss-optimized Switch Engine exhibiting a total of 2.3 ± 0.2 dB loss per user, 1.8 dB from the WDM and 0.5 dB from the LLOS. Each user module included a polarization controller (PC) for the polarization rotation of each photon pair and a polarization analysis module (PAM) to separate the two orthogonal polarization bases (HV and DA). Single photon detection was achieved by 16 free-running high-efficiency (10 Hz dark count rate at ~90% efficiency, and <10 ps jitter) superconducting nanowire single-photon detectors (SNSPDs), each preceded by a PC to maximize the detection efficiency, with one detector assigned to each of the four polarization states per user. Detection pulses were registered by a 16-channel time tagging module (Time Tagger Ultra by Swabian Technologies) with an RMS jitter at 8 ps. Because all SNSPDs were co-located in a single cryostat and synchronized by the same time tagger, the four users were logically separated at the protocol level. We then implemented a QKD software generation module using the BBM92 protocol for raw key generation, followed by the Cascade protocol for error correction and Fast Fourier Transform-based privacy amplification. The software was able to produce real-time secure keys that were stored on dedicated per-user QKD provider buffers. At the key management layer, the in-house SD-KMS continuously harvested 256-bit keys via periodic ETSI GS QKD 014 requests [10]. Each KMS node instance features a local KMS agent that handles data-plane cryptographic operations, while a centralized SD-KMS controller provides network-wide monitoring and management. This controller also governs the switching operations of the LLOS and dynamically configures active QKD links according to the current entanglement distribution pairs and incoming secret key requests. The stored QKD keys for each user pair allowed the KMS agents to provision keys via the ETSI 014 interface to FortiGate NGFWs, which established QKD-encrypted IPsec tunnels.

## 3. Results

We experimentally evaluated the deployed four-user actively reconfigurable network over one hour, assessing quantum-link performance using the actual secret-key rates (SKRs) obtained through post-processing of the detection data, and integration with the SD-KMS. QKD performance depends on the temporal alignment of the detection events between the two users of each link and on Quantum Bit Error Rate (QBER) minimization. To achieve this, an initialization procedure was performed for each link at startup and after every switching event. This procedure includes the calculation of the relative delay between the detection streams to synchronize the detection events and polarization compensation to minimize QBER. Following initialization, secure-key generation begins and the resulting keys are served to the overarching KMS. The KMS controller continuously monitors the key buffers occupancy and links health status and provisions the secure keys to the external encryptors. Fig. 2a) shows the mean coincidence rate as a function of the relative delay during runtime, characterizing the temporal alignment of the detection events and the source performance. Mean coincidence values ranged from 10k–17k per second while low detector jitter combined with the short fiber links enabled the use of narrow coincidence windows ranging from 100 to 150 ps. Regarding the QKD performance, QBER values after the initialization procedure were kept low, predominantly between 0.5% and 3% with the corresponding SKR of 4060 to 7081 bps, as shown in Fig. 2b) and c), respectively. SKR and fidelity mean values are presented for all the links in Table 1. Each user held a QKD provider buffer with a capacity of 10,000

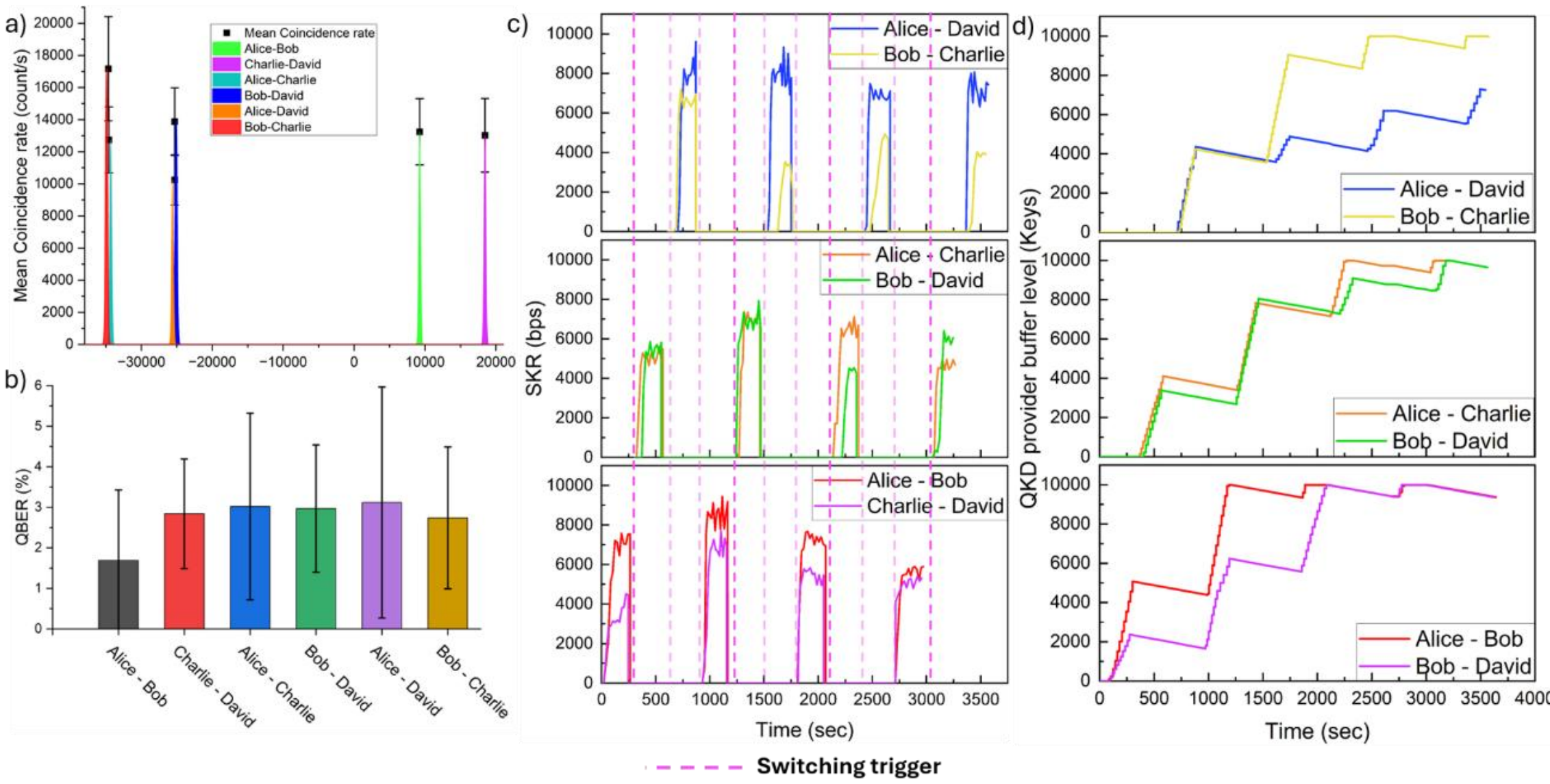

Fig. 2 (a) Coincidence rate as a function of relative detection delay, (b) mean QBER for each user pair, (c) SKR over time for the six user pairs, and (d) QKD key buffer occupancy over time for each user pair.

256-bit keys, which was continuously fed the KMS and was replenished during the active link key-generation periods, as depicted in Fig. 2d). The interface of each node remained active even when key generation was disabled, allowing the KMS agents to continuously fetch QKD keys at a rate of 1 key/s. Simultaneously, the encryptors requested keys at a rate of 0.5 key/min per node to support real-time quantum-safe applications. At each layer, since the key consumption rate was lower than the key-generation rate during each link's active cycle, all key buffers accumulated a surplus of key material and approached their maximum capacity towards the end of the experiment.

Table 1. Average SKR and fidelity with their corresponding standard deviation for all the links

| | Alice - Bob | Charlie - David | Alice - Charlie | Bob - David | Alice - David | Bob - Charlie |
|---|---|---|---|---|---|---|
| Channels | C33 – C35 | C32 – C36 | C33 – C35 | C32 – C36 | C33 – C35 | C32 – C36 |
| SKR (bps) | 6583 ($\pm$1893) | 5157 ($\pm$1306) | 5500 ($\pm$1324) | 5241 ($\pm$1850) | 7081 ($\pm$1671) | 4060 ($\pm$2032) |
| Fidelity | 0.97 ($\pm$0.03) | 0.96 ($\pm$0.02) | 0.95 ($\pm$0.03) | 0.96 ($\pm$0.02) | 0.95 ($\pm$0.04) | 0.96 ($\pm$0.03) |

## 4. Conclusion

This work demonstrated a dynamic 4-node entanglement-based actively switched QKD network supporting any to any communication and a SD-KMS, enabling secure and dynamic reconfiguration of quantum links. The experimental results validated the successful operation of automated switching and secure key generation between all the network users. The integration of SD-KMS provided an abstracted centralized network-wide orchestration with the capability of accommodating secure keys from different QKD systems and technologies, while it seamlessly supplied them to the SAEs. Overall, the proposed architecture showcases a scalable, multi-user and quantum-secure approach for large-scale dynamic entanglement-based QKD network deployments.

## Acknowledgements

This work was funded by QOSiLICIOUS (GA 101187539) and 6G-CATS (GA No.101287162).